\PassOptionsToPackage{table}{xcolor}
\documentclass[10pt,conference]{IEEEtran}

\usepackage{saner20}
\usepackage{subcaption}

\begin{document}

\title{Cross-Dataset Design Discussion Mining}

\author{\IEEEauthorblockN{Alvi Mahadi}
\IEEEauthorblockA{\textit{Department of Computer Science} \\
\textit{University of Victoria, Canada}\\
alvi.utsab@gmail.com}
\and
\IEEEauthorblockN{Karan Tongay}
\IEEEauthorblockA{\textit{Department of Computer Science} \\
\textit{University of Victoria, Canada}\\
karantongay@gmail.com}
\and
\IEEEauthorblockN{Neil A. Ernst}
\IEEEauthorblockA{\textit{Department of Computer Science} \\
\textit{University of Victoria, Canada}\\
neil@neilernst.net}
}

\maketitle

\begin{abstract}
Being able to identify software discussions that are primarily about design---which we call design mining---can improve documentation and maintenance of software systems. Existing design mining approaches have good classification performance using natural language processing (NLP) techniques, but the conclusion stability of these approaches is generally poor. A classifier trained on a given dataset of software projects has so far not worked well on different artifacts or different datasets. In this study, we replicate and synthesize these earlier results in a meta-analysis. We then apply recent work in transfer learning for NLP to the problem of design mining. However, for our datasets, these deep transfer learning classifiers  perform no better than less complex classifiers. We conclude by discussing some reasons behind the transfer learning approach to design mining. 
\end{abstract}


\noindent\textbf{The replication package for this paper may be found at \url{https://doi.org/10.5281/zenodo.3590126}}
\section{Introduction}
\label{sect:introduction}

Design discussions are an important part of software development. Software design is a highly interactive process and many decisions involve considerable back and forth discussion. These decisions greatly impact software architecture \cite{Kazman2016,woods16}.
However, software design is a notoriously subjective concept. For the related term `software architecture', for example, the Software Engineering Institute maintains a list of over 50 different definitions \cite{Software-Engineering-Institute:2010aa}). This subjectivity makes analyzing design decisions difficult \cite{Shahbazian2018}. Researchers have looked for ways in which design discussions could be automatically extracted from different types of software artifacts \cite{Brunet2014,Shakiba2016,Alkadhi2017,Motta2018,Zanaty2018,Viviani2018b,Viviani2019}. This automatic extraction, which we call \emph{design mining}, is a subset of research based on mining software repositories. The potential practical relevance of research on design mining include supporting design activities, improving traceability to requirements, enabling refactoring, and automating documentation.

A design mining study uses a corpus consisting of \emph{discussions}, in the form of software artifacts like pull requests, and manually labels those discussions targeting design topics, according to a coding guide. Machine learning classifiers such as support vector machines learn the feature space based on a vectorization of the discussion, and are evaluated using the gold set with metrics like area under the ROC curve.
Producing a good-performing, validated classifier has been the main objective to date for design mining research. Apart from a validation study in Viviani et al. \cite{Viviani2019}, however, the practical relevance (cf. \cite{Oliveira-Neto:2019aa}) of design mining has not been studied in detail. Practical relevance in the context of design mining means a classifier with broad applicability to different design discussions, across software projects and across artifact types. Our goal is a practically relevant classifier, which we could run on a randomly selected Github project and identify with high accuracy that project's design discussions. 

Practical relevance can be seen as a form of external validity or generalizability, and the underlying concept is \emph{conclusion stability}, after Menzies and Sheppherd \cite{Menzies2012}. 
Design mining research has to date performed poorly when applied to new datasets (which we elaborate on in a discussion of related work, following). 
This is problematic because positive, significant results in a single study are only of value if they lend confidence to scientific conclusions. 
Conclusion stability, and a researcher's confidence they have found a true effect, relies on the ability to transfer a learner from an initial dataset (given the choices above) to other datasets, which may or may not match the initial study conditions.
To gain more insight on the challenges of conclusion stability in design mining, we report on two research objectives.

\textbf{RO1}: Our first research objective is to assess, by replication, whether it is possible to accurately label a given natural language discussion as pertaining to software design. We conduct an exact replication of the study of Brunet et al. \cite{Brunet2014}. We also examined, but did not exactly replicate, similar studies (as shown in Table \ref{table:related}).
This allows us to identify similar, replicable elements in the studies: common research goals (i.e., identifying design), additional datasets and common research protocols (such as the removal of stopwords). We also modernize the analysis approaches (e.g., by incorporating stratification).
By using datasets from existing studies, but varying the research protocols, we are conducting \emph{operational} replications (using the terminology of G\'{o}mez et al. \cite{Gmez2014}). This allows us to characterize each study's choices.
The operational replication also helps to explain the study's internal validity, which in turn should provide a richer basis for claims of external validity \cite{Siegmund2015}.

\textbf{RO2:} To what extent can we transfer classifiers trained on one data set to other data sets? 
Are there ways to improve conclusion stability for a design mining classifier on an entirely different software discussion dataset? 
We examine new advances in natural language processing (NLP) which focus on improving transfer learning between projects. 

To tackle the first objective, we conduct an operational replication of the pioneering design mining work of Brunet et al. \cite{Brunet2014}. 
We first replicate, as closely as possible, the study Brunet et al. conducted. 
We then try to improve on their results in design mining with new analysis approaches.
We demonstrate improved ways to detect design discussions using word embeddings and document vectors. 

For the second objective, we report on cross-project transfer of a classifier. We begin with the approaches described in \textbf{RO1}, and apply them to new, out of sample datasets. We then apply a recently introduced deep transfer learning approach called ULMFiT \cite{Howard:2018}. We characterize the different datasets we study to understand what commonalities and differences they exhibit. 
Our contributions:
\begin{itemize}
  \item Improved classification results using word embeddings and stratification, with AUC of 0.84.
  \item Meta-analysis using vote-counting of previous studies. 
  \item StackOverflow design-related dataset, with document vectors.
  \item Characterization of the conclusion stability of NLP models for design mining.
  \item Explanation of ULMFiT approach to inductive transfer learning on design discussions.
\end{itemize}

\section{Background and Related Work}
\label{sect:related_work}

Our paper brings together two streams of previous research. First,  we highlight work on transfer learning in software engineering.
Secondly, we discuss previous work in mining design discussions and summarize existing results as an informal meta-analysis.

\subsection{Cross-Project Classifiers in Software Engineering}
A practically relevant classifier is one that can ingest a text snippet---design discussion---from a previously unseen software design artifact, and label it \textsf{Design/Not-Design} with high accuracy. 
Since the classifier is almost certainly trained on a different set of data, the ability to make cross-dataset classifications is vital. 
Cross-dataset classification \cite{Zimmermann:2009} is the ability to train a model on one dataset and have it correctly classify other, different datasets. 

The challenge is that the underlying feature space and distribution of the new datasets differ from that of the original dataset, and therefore the classifier often performs poorly. 
For software data, the differences might be in the type of software being built, the size of the project, or how developers report bugs.
To enable cross-domain learning without re-training the underlying models, the field of transfer learning applies machine learning techniques to improve the transfer between feature spaces \cite{Pan:2010}. Typically this means learning the two feature spaces and creating mapping functions to identify commonalities.

A related concept is the notion of \emph{conclusion stability} from Menzies and Sheppherd \cite{Menzies2012}. 
Conclusion stability is the notion that an effect X that is detected in one situation (or dataset) will also appear in other situations. 
Conclusion stability suggests that the theory that predicts an effect X holds (transfers) to other datasets. In design mining, then, conclusion stability is closely tied to the ability to transfer models to different datasets.

There have been several lines of research into transfer learning in software engineering. We summarize a few here. 
Zimmermann et al. \cite{Zimmermann:2009} conducted an extensive study of conclusion stability in defect predictors. 
Their study sought to understand how well a predictor trained with (for example) defect data from one brand of web browser might work on a distinct dataset from a competing web browser. 
Only 3.4\% of cross-project predictions achieved over 75\% accuracy, suggesting transfer of defect predictors was difficult. 

Following this work, a number of other papers have looked at transfer learning within the fields of effort estimation and defect prediction. Sharma et al \cite{Sharma:2019aa} have applied transfer learning to the problem of code smell detection. 
They used deep learning models and showed some success in transferring the classifier between C\# and Java. However, they focus on source code mining, and not natural language discussions. Code smells, defect prediction, or effort estimation are quite distinct from our work in design discussion, however, since they tend to deal with numeric data, as opposed to natural language.

Other approaches include the use of bellwethers \cite{Krishna:2016:TMA:2970276.2970339}, exemplar datasets that can be used as simple baseline dataset for generating quick predictions. The concept of bellwether for design is intriguing, since elements of software design, such as patterns and tactics, are generalizable to many different contexts.

As far as we know, there has not been any use of transfer learning in natural language processing tasks for software engineering beyond the early results of Robbes and Janes \cite{Robbes:2019}, who reported on using ULMFiT \cite{Howard:2018} for sentiment analysis. We also use the transfer NLP potential of ULMFiT, which we discuss in \S \ref{sec:ulmfit}. Robbes and Janes emphasized the importance of pre-training the learner on (potentially small) task-specific datasets. We extensively investigate the usefulness of this approach with respect to design mining.

\subsection{Mining Design Discussions}
\label{sec:related_mining}
While repository mining of software artifacts has existed for two decades or more, mining repositories for \emph{design-related} information is relatively recent. In 2011 Hindle et al. proposed labeling non-functional requirements in order to track a project's relative focus on particular design-related software qualities, such as maintainability \cite{hindle11msr}. Hindle later extended that work \cite{HindleBZN15} by seeking to cross-reference commits with design documents at Microsoft. Brunet et al. \cite{Brunet2014} conducted an empirical study of design discussions, and is the target of our strict replication effort. They pioneered the classification approach to design mining: supervised learning by labeling a corpus of design discussions, then training a machine learning algorithm validated using cross-validation. 

In Table \ref{table:related} we review some of the different approaches to the problem, and characterize them along the dimensions of how the study defined ``design'', how prevalent design discussions were, what projects were studied, and overall accuracy for the chosen approaches. We then conduct a rudimentary vote-counting meta-review \cite{Pickard1998} to derive some overall estimates for the feasibility of this approach (final row).

\begin{table*}
    \centering
        \caption{\small{Comparison of recent approaches to design discussion detection. Effectiveness captures the metric the paper reports for classifier effectiveness (accuracy, precision, recall, F1). NB: Naive Bayes; LR: Logistic Regression; DT: Decision Tree; RF: Random Forest; SVM: Support Vector Machine}} 
        \label{table:related}
    \begin{tabular}{cp{1.4cm}p{1.4cm}p{1.2cm}p{1.4cm}p{1.2cm}p{2.7cm}p{2.4cm}}
        \toprule
    \textbf{Study} &  \textbf{Projects Studied} & \textbf{Data Size} & \textbf{ML Algorithm} & \textbf{Effectiveness} & \textbf{Prevalence} & \textbf{Defn. of Design} & \textbf{Defn. of Discussion}\\ 
    \midrule
    Brunet \cite{Brunet2014} & 77 high importance Github projects & 102,122 comments  & NB DT & \textbf{Acc: 0.86/0.94} & 25\% of discussions & design is the process of discussing the structure of the code to organize abstractions and their relationships. & A set of comments on pull requests, commits, or issues\\ 
    Alkadhi17 \cite{Alkadhi2017}  & 3 teams of undergrads & 8,702 chat messages of three development teams & NB SVM + undersampling & \textbf{Prec: 0.85} & 9\% of messages & Rationale captures the reasons behind decisions. & Messages in Atlassian HipChat\\ 
    Alkadhi18 \cite{Alkadhi2018} & 3 Github IRC logs & 7500 labeled IRC messages & NB SVM & \textbf{Prec. 0.79} & 25\% of subset labeled & Rationale captures the reasons behind decisions. & IRC logs from Mozilla\\ 
    Zanaty \cite{Zanaty2018}  & OpenStack Nova and Neutron & 2817 comments from 220 discussions & NB  SVM KNN DT & \textbf{Prec: 0.66 Recall: 0.78} & 9-14\% & Brunet's \cite{Brunet2014} & Comments on code review discussions\\ 
    Shakiba \cite{Shakiba2016}  & 5 random Github/SF & 2000 commits & DT  RF NV  KNN & \textbf{Acc: 0.85} & 14\% of commits & None. & Commit comments\\ 
    Motta \cite{Motta2018}  & KDELibs & 42117 commits, 232 arch & Wordbag matching & n/a & 0.6\% of commits & Arch keywords from survey of experts & Commit comments\\ 
    Maldonado \cite{Maldonado2017} & 10 OSS Projects & 62,566 comments & Max Entropy & \textbf{F1: 0.403 } & 4\% design debt & Design Debt: comments indicate that there is a problem with the design of the code & Code comment\\ 
    Viviani18 \cite{Viviani2018} & Node, Rust, Rails & 2378 design-related paragraphs & n/a (qualitative) & n/a & 22\% of paragraphs & A piece of a discussion relating to a decision about a software system’s design that a software development team needs to make & Paragraph, inside a comment in a PR\\
    Viviani19 \cite{Viviani2019} & Node, Rust, Rails & 10,790 paragraphs from 34 pull requests & RF & \textbf{AUC 0.87} & 10.5\% of paragraphs & same as Viviani18 & same as Viviani18\\
    (This paper) & StackOverflow discussions & 51,990 questions and answers & LR/ SVM/ ULMFiT & \textbf{AUC: 0.84} & n/a  & A question or answer with the tag ``design" & StackOverflow question/answer \\      \midrule
    Meta & Open-source & 29,298 & n/a &\textbf{Acc: 0.85-0.94} & 13.78\% & n/a & n/a \\         \bottomrule
\end{tabular}
\end{table*}
    
\noindent\textbf{Defining Design Discussions}---The typical unit of analysis in these design mining studies is the ``discussion'', i.e., the interactive back-and-forth between project developers, stakeholders, and users. As Table \ref{table:related} shows, this changes based on the dataset being studied. A discussion can be code comments, commit comments, IRC or messaging application chats, Github pull request comments, and so on. The challenge is that the nature of the conversation changes based on the medium used; one might reasonably expect different discussions to be conducted over IRC vs a pull request.

\noindent\textbf{Frequency of Design Discussions}---Aranda and Venolia \cite{Aranda_2009} pointed out in 2009 that many software artifacts do not contain the entirety of important information for a given research question (in their case, bug reports). Design is, if anything, even less likely to appear in artifacts such as issue trackers, since it operates at a higher level of abstraction. Therefore we report on the average prevalence of design-related information in the studies we consider. On average 14\% of a corpus is design-related, but this is highly dependent on the artifact source. 
    
\noindent\textbf{Validation Approaches for Supervised Learning}---In Table \ref{table:related} column \textsf{Effectiveness} reports on how each study evaluated the performance of the machine learning choices made. These were mostly the typical machine learning measures: accuracy (number of true positives + true negatives divided by the total size of the labeled data), precision and recall (true positives found in all results, proportion of results that were true positives), and F1 measure (harmonic mean of precision and recall). 
Few studies use more robust analyses such as AUC (area under ROC curve, also known as balanced accuracy, defined as the rate of change). 
Since we are more interested in design discussions, the minority class of the dataset, AUC or balanced accuracy gives a better understanding of the result, because of the unbalanced nature of the dataset.

\noindent\textbf{Qualitative Analysis}
The qualitative approach to design mining is to conduct what amount to targeted, qualitative assessments of projects. The datasets are notably smaller, in order to scale to the number of analysts, but the potential information is richer, since a trained eye is cast over the terms. The distinction with supervised labeling is that these studies are often opportunistic, as the analyst follows potentially interesting tangents (e.g., via issue hyperlinks). Ernst and Murphy \cite{Ernst:2012wf} used this case study approach to analyze how requirements and design were discussed in open-source projects. One follow-up to this work is that of Viviani, \cite{Viviani2018,Viviani2019}, papers which focus on rubrics for identifying design discussions. The advantage to the qualitative approach is that it can use more nuance in labeling design discussions at more specific level; the tradeoff of course is such labeling is labour-intensive. 

\noindent\textbf{Summary}
A true meta-analysis \cite{Pickard1998, Kitchenham2019} of the related work is not feasible in the area of design mining. 
Conventional meta-analysis is applied on primary studies that conduct experiments in order to support inference to a population, which is not the study logic of the studies considered here. 
For example, there are no sampling frames or effect size calculations. 
One approach to assessing whether design mining studies have shown the ability to detect design is with vote-counting (\cite{Pickard1998}), i.e., count the studies with positive and negative effects past some threshold. 

As a form of vote-counting, the last row of Table \ref{table:related} averages the study results to derive estimates. 
On average, each study targets 29,298 discussions for training, focus mostly on open-source projects, and find design discussions in 14\% of the discussions studied. 
As for effectiveness of the machine learning approaches, here we need to define what an `effective' ML approach is. 
For our purposes, we can objectively define this as ``outperforms a baseline ZeroR learner". 
The ZeroR learner labels a discussion with the majority class, which is typically ``non-design". 
In a balanced, two label dataset, the ZeroR learner would therefore achieve accuracy of 50\%. 
In an unbalanced dataset, which is the case for nearly all design mining studies, ZeroR is far more `effective'. Using our overall average of 14\% prevalence, a ZeroR learner would achieve accuracy of $(1-0.14) = 0.86$. 
This is the baseline for accuracy effectiveness. 
For precision and recall, ZeroR would achieve 0.86 and 1, for an F1 score of 0.93. 
Comparing this baseline to the studies above, we find that only Brunet and our approach below surpass this baseline. In other words, few studies are able to supersede random, majority-class labeling.

\section{Research Objective 1: Design Mining Replication}

\subsection{Strict Replication}
\label{sect:strict_replication}

We now turn to RO1, replicating the existing design mining studies and exploring the best combination of features for state of the art results. 
To begin, we conduct a strict replication (after Gómez et al. \cite{Gmez2014}), a replication with little to no variance from the original study, apart from a change in the experimenters. 
However, given this is a computational data study, researcher bias is less of a concern than lab or field studies (cf. \cite{Storey:2019aa}).
The purpose of these strict replications is to explain the current approaches and examine if recent improvements in NLP might improve the state of the art.

\begin{figure}[hbt]
\centering
  \includegraphics[width=0.45\textwidth]{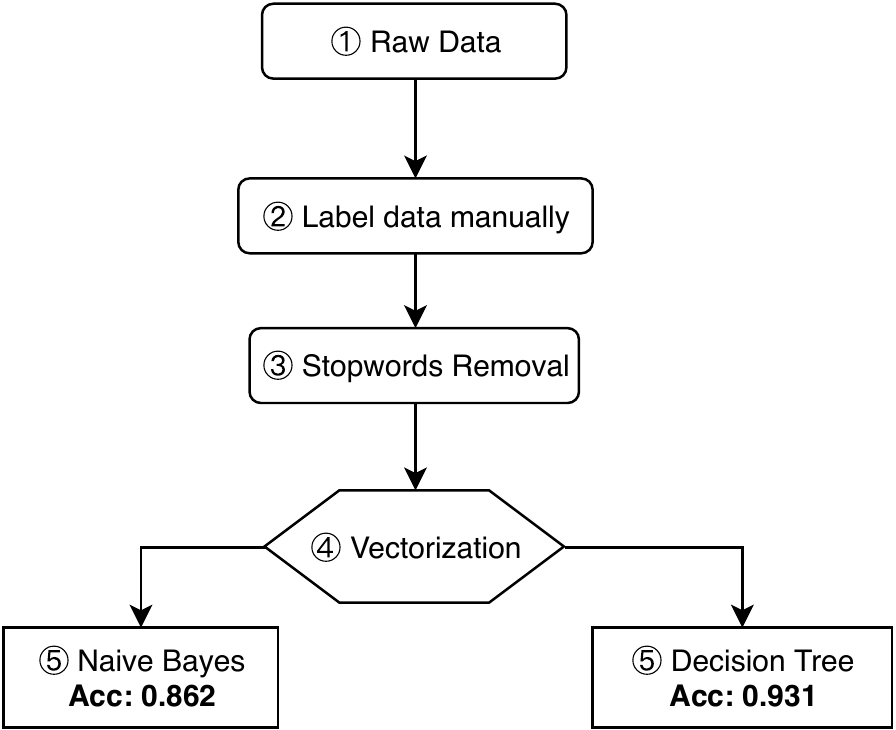}
  \caption{Protocol map of Brunet \cite{Brunet2014} study}
  \label{fig:brunet-protocol}
\end{figure}

To explain the differences in studies, we use protocol maps, a graphical framework for explaining an analysis protocol. This graphical representation is intended to provide a visual device for comprehending the scope of analysis choices in a given study. Fig. \ref{fig:brunet-protocol} shows a protocol map for the strict replication. 
The enumerated list that follows matches the numbers in the protocol diagram. 
\begin{enumerate}
  \item Brunet's study \cite{Brunet2014} selected data from 77 Github projects using their discussions found in pull requests and issues.
  \item Brunet and his colleagues labeled 1000 of those discussions using a coding guide.
  \item Stopwords were removed. They used NLTK stopwords dictionary and self defined stopsets.
  \item The data were vectorized, using a combined bigram word feature and using the NLTK BigramCollectionFinder to take top 200 ngrams.
  \item Finally, Brunet applied two machine learning approaches, Naive Bayes and Decision Trees. 10-fold cross validation produced the results shown in Fig. \ref{fig:brunet-protocol}: mean accuracy of \textbf{0.862} for NaiveBayes, and \textbf{0.931} for Decision Trees, which also several orders of magnitude slower.
\end{enumerate}

We followed this protocol strictly. We downloaded the data that Brunet has made available; applied his list of stop words; and then used Decision Trees and NaiveBayes to obtain the same accuracy scores as his paper. The only difference is the use of \textsf{scikit-learn} for the classifiers, instead of NLTK. Doing this allowed us to match the results that the original paper \cite{Brunet2014} obtained.

We did notice one potential omission. 1000 sentences are manually classified in Brunet's dataset \cite{Brunet2014}. However, only 224 of them are design, which indicates serious imbalance in the data. As a result, the accuracy measure, which assumes a balanced set of classes, likely overstates the true validity of this approach.

\begin{figure*}[hbt]
\centering
  \includegraphics[width=0.9\textwidth]{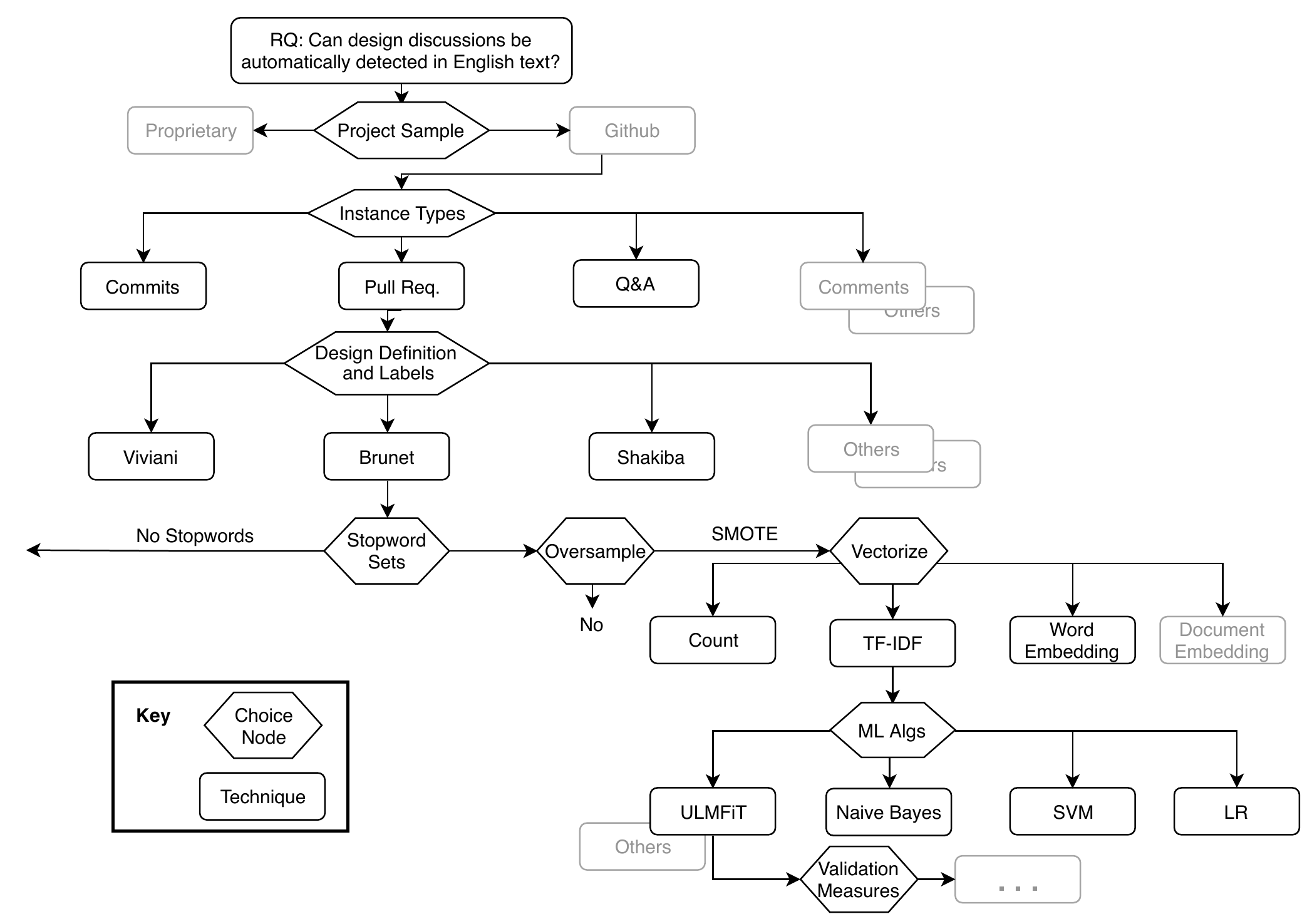}
  \caption{Protocol map of possible research paths for design mining studies.}
  \label{fig:extensions}
\end{figure*}

\subsection{Extending the Replication}
\label{sect:extending_the_replication}

A strict replication is useful to confirm results, which we did, but does not offer much in the way of new insights into the underlying research questions. In this case, we want to understand how to best extract these design discussions from \emph{any} corpora. This should help understand what features are important for our goal of improving conclusion stability.

Shepperd \cite{Shepperd2018} shows that focusing (only) on replication ignores the real goal, namely, to increase confidence in the result. Shepperd's paper focused on the case of null-hypothesis testing, e.g., comparison of means. In the design mining problem, our confidence is based on the validation measures, and we say (as do Brunet and the papers we discussed in \S \ref{sec:related_mining}) that we have more confidence in the result of a classifier study if the accuracy (or similar measures of classifier performance) is higher. 

However, this is a narrow definition of confidence; ultimately we have a more stable set of conclusions (i.e. that design discussions can be extracted with supervised learning) if we can repeat this study with entirely different datasets. We first discuss how to improve the protocol for replication, and then, in Section \ref{sect:conclusion_s}, discuss how this protocol might be applied to other, different datasets.

We extend the previous replication in several directions. Fig. \ref{fig:extensions} shows the summary of the extensions, with many branches of the tree omitted for space reasons.
One immediate observation is that it is unsurprising conclusion stability is challenging to achieve, given the vast number of analysis choices a researcher could pursue.
We found several steps where Brunet's original approach could be improved. These improvements also largely apply to other studies shown in Table \ref{table:related}. 

We switched to use balanced accuracy, or area under the receiver operating characteristic curve (AUC-ROC or AUC), since it is a better predictor of performance in imbalanced datasets\footnote{defined in the two-label case as the True Positive Rate + the False Positive Rate, divided by two}.

\begin{figure}[hbt]
\centering
  \includegraphics[width=0.35\textwidth]{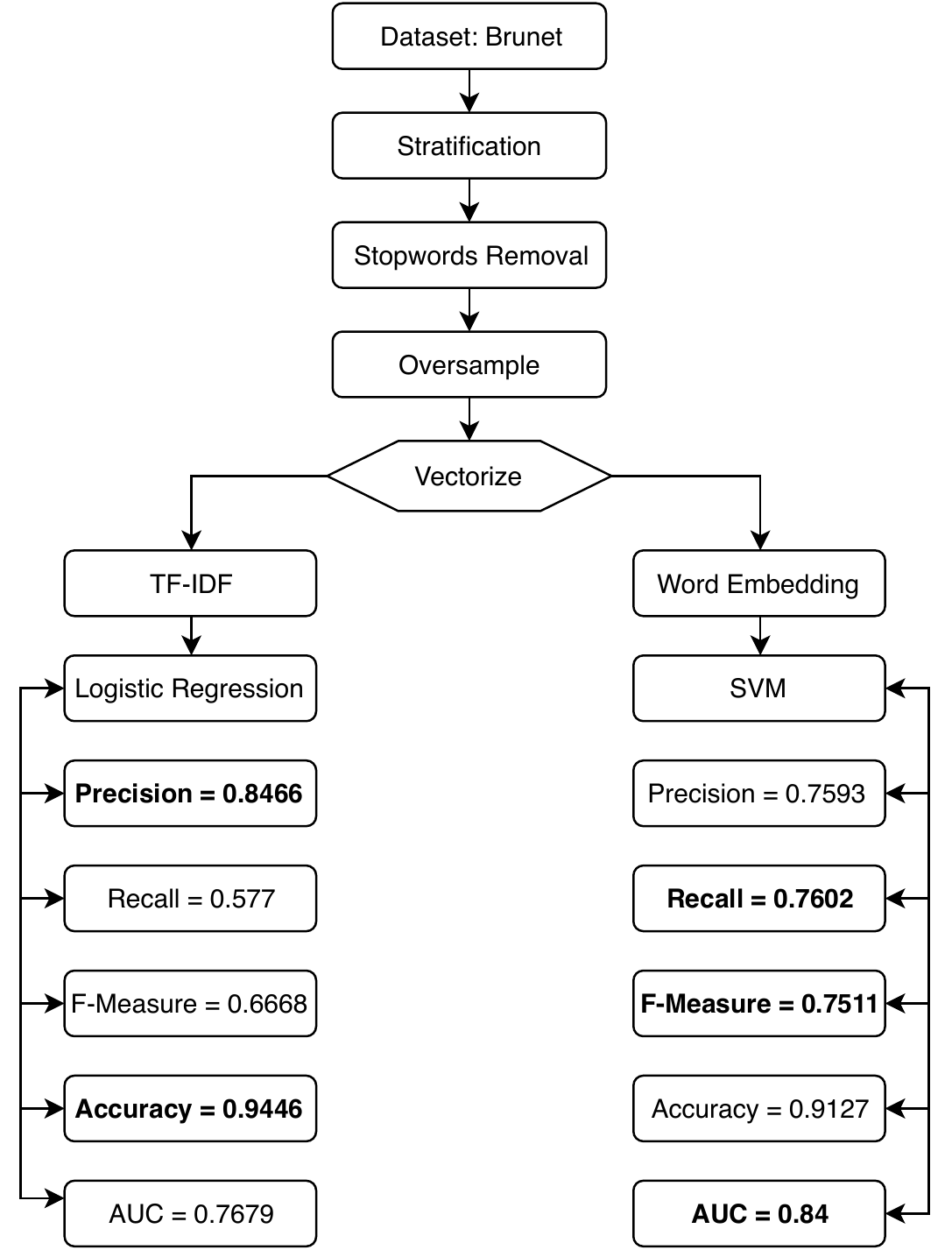}
  \caption{Preferred Design Mining Method \textsf{NewBest}. Numbers are mean of 10-fold cross validation.}
  \label{fig:preferred_method}
\end{figure}

\subsubsection{Vectorization Choices}
\label{sec:wordembed}
Vectorization refers to the way in which the natural language words in a design discussion are represented as numerical vectors, which is necessary for classification algorithms. We present four choices: one, a simple count; two, term-frequency/inverse document frequency (TF-IDF), three, word embeddings, and four, document embeddings. The first two are relatively common so we focus on the last two.

Word embeddings are vector space representations of word similarity. Our intuition is this model should capture design discussions better than other vectorization approaches. A word embedding is first trained on a corpus. In this study, we consider two vectorization approaches, and one similarity embedding. ``Wiki'' is a Fasttext embedding produced from training on the Wikipedia database plus news articles \cite{mikolov2018advances}, and GloVe, trained on web crawling \cite{pennington2014glove}. The final embedding is trained on the StackOverflow dataset, courtesy of Efstathiou et al. \cite{Efstathiou:2018aa}. While Wikipedia considers more words in English, the StackOverflow dataset should be more representative of the software domain. The embedding is then used to either a) train a classifier like Logistic Regression by passing new discussions to the embedding, and receiving a vector of its spatial representation in return; b) expanding the scope of small discussions by adding related words to the sentence (StackOverflow).

As Fig. \ref{fig:extensions} shows, there are several ways in which vectorization applies. We also wanted to see if we could expand the size of the training set by using the in-built capability of a word embedding to identify similar words (i.e., words that are close in vector space). The intuition is that since the discussions from the Brunet dataset are typically quite short, we could add similar words  to extend the vocabulary. However, the vocabulary expansion approach did not make any difference to our accuracy results, likely because domain-specific terms like ``library" were overwhelmed with standard English terms like ``thus|hence''.

\subsubsection{Document Vectors and the StackOverflow Design Corpus}
\label{sec:doc2vec}
Extending word vectors, we can also capture the spatial representation of entire discussions. 
Document vectors, introduced in \cite{doc2vec14}, are extensions to word embeddings that add an extra dimension to the vector space to capture the document of origin (in this case, a design discussion). 
The approach has been shown to improve the ability to capture discussion-wide meaning, where a word embedding approach alone would be focused on a smaller window of words. 
We used the Gensim Doc2Vec class to train a document embedding on 26,969 StackOverflow questions and answers, that were tagged with the label `design' (which we extracted from the SOTorrent dataset of Baltes et al. \cite{BaltesDT008}), combined with 25,000 random questions \emph{not} tagged design.\footnote{This dataset can be found as part of our replication package}

We processed the data to remove stopwords, HTML and \verb| <code>| tags (including the code snippets found within). 
We also removed graphical or web design discussions, where they had tags that co-occurred with the `design' tag, such as \texttt{CSS,HTML}. 
We then trained a document vector based on the 51,969 documents in the corpus, and used logistic regression to classify the documents as either design or not. 
Results are good on the internal, within-sample classification, even for the test set: accuracy of 0.934 in training, and 0.932 for test (held-out) data. This dataset is balanced so accuracy is a reasonable validation metric.

\subsubsection{Other Extensions}

We used imbalance correction in order to account for the fact design discussion make up only 14\% (average) labels. We took two approaches.
One, we stratified folds to keep the ratio of positive and negative data equal. 
After stratifying, we have again run the experiment described in \cite{Brunet2014} and examined that the accuracy dropped significantly from reported 94\% to around 87.6\% where our experiment achieved an accuracy of around 94\%. 
We use SMOTE \cite{Chawla:2002aa} to correct for imbalanced classes in train data. Recall from Table \ref{table:related} that design discussion prevalence is at best 14\%. This means that training examples are heavily weighted to non-design instances. As in \cite{Alkadhi2017}, we correct for this by increasing the ratio of training instances to balance the design and non-design instances. We have oversampled the minority class (i.e., `design'). 

We also hypothesized that the software-specific nature of design discussions might mean using non-software training data would not yield good results. Specifically, when it comes to stopword removal, we used our own domain-specific stopword set along with the predefined English stopwords (of scikit-learn). We also searched for other words that may not mean anything significant, such as `lgtm' (`looks good to me') or `pinging', which is a way to tag someone to a discussion. These stopwords may vary depending on the project culture and interaction style, so we removed them. 

\subsubsection{Best Performing Protocol}
After applying these extensions, Fig. \ref{fig:preferred_method} shows the final approach. Ultimately, for our best set of choices we were able to obtain an AUC measure of 0.84, comparable to the unbalanced accuracy Brunet reported of 0.931.

Logistic Regression with TF-IDF vectorization gives the best results in terms of Precision and Accuracy. 
On the other hand, Word Embedding with Support Vector Machine provides best results in terms of Recall, F-Measure and Balanced Accuracy or AUC. 
Since we are interested in the `design' class which is the minority class of the dataset, highest Recall value should be more acceptable than Precision. 
As a result we created a \textsf{NewBest} classifier based on the combination  of `Word Embedding' and `Support Vector Machine' (right hand of Fig. \ref{fig:preferred_method}).
\begin{table*}[h]
 \centering
 \caption{Datasets used for within and cross-dataset classification. All datasets are English-language}
 \label{tbl:newdatasets}
 \begin{tabular}{ccp{2.4cm}p{1cm}p{1cm}p{2.4cm}p{1.4cm}p{1.4cm}} 
 \toprule
 \textbf{Citation} & \textbf{Dataset} & \textbf{Type}  & \textbf{Total instances} & \textbf{Design instances} & \textbf{Projects} & \textbf{Mean Discussion Length (words) } & \textbf{Vocabulary Size (words)} \\
 \midrule
 \cite{Brunet2014} & Brunet 2014  &  Pull requests & 1,000 & 224 & BitCoin, Akka, OpenFramework, Mono, Finagle & 16.97 & 3,215 \\
\cite{Shakiba2016} & Shakiba 2016  & Commit messages & 2,000 & 279 & Random Github and SourceForge  & 7.43 & 4,695 \\
 \cite{Viviani2018} & Viviani 2018   &   Pull requests & 5,062 & 2,372 & Node, Rust, Rails & 36.13 &  24,141 \\
\cite{Maldonado2017} & SATD  & Code comments & 62,276 & 2,703 & 10 Java incl Ant, jEdit, ArgoUML & 59.13 & 49,945 \\ self & StackOverflow & StackOverflow questions & 51,989 & 26,989 & n/a & 114.79 & 252,565 \\
 \bottomrule
 \end{tabular}
 \end{table*}

 \section{Research Objective 2: Conclusion Stability}
\label{sect:conclusion_s}
In this section we build on the replication results and enhancements of our first research objective. We have a highly accurate classifier, NewBest, that does well \emph{within-dataset}. We now explore its validity when applied to other datasets, i.e., whether it has conclusion stability. 

In \cite{Menzies2012}, Menzies and Shepperd discuss how to ensure conclusion stability. They point out that predictor performance can change dramatically depending on dataset (as it did in Zimmermann et al. \cite{Zimmermann:2009}). Menzies and Shepperd specifically analyze prediction studies, but we believe this can be generalized to classification as well. Their recommendations are to a) gather more datasets and b) use train/test sampling (that is, test the tool on different data entirely). 

In this section we evaluate a classifier trained on one dataset to a different dataset, but consisting of the same types of discussions. Before beginning to apply learners to different datasets, it makes sense to ask if this transfer is reasonable. For example, in Zimmermann et al. \cite{Zimmermann:2009} the specific characteristics of each project were presented in order to explain the intuition behind transfer. 
E.g., should a discussion of design in StackOverflow be transferable, that is, considered largely similar to, one from Github pull requests?

\subsection{Research Method}
In Table \ref{tbl:newdatasets} we illustrate each of the datasets considered in this paper. In Table \ref{tbl:examples} we show some sample design discussions from each.
Since performance of transfer learning is largely based on similarity between projects (i.e., feature spaces), we would expect to see better AUC results for cross-project prediction if data sources are the same (e.g. pull requests), projects are the same, and/or the platforms are the same (e.g. Github). 

We test the ability to transfer classifiers to new types of discussions and datasets.
We applied the best protocol result from above. That is, the NewBest classifier, using \textsf{stopwords+oversampling+TF-IDF+Logistic Regression}. 
We train this classifier on the Brunet \cite{Brunet2014} data, and the other 4 datasets described in Table \ref{tbl:newdatasets}.

We then apply the trained model, as well as the ULMFiT model described below, to each dataset in turn (thus, 5 comparisons, including within-project labeling for a baseline).

 \begin{table}[htb]
 \centering
 \caption{Sample (raw) design discussions, pre data cleaning.}
 \label{tbl:examples}
 \begin{tabular}{cp{6.2cm}} 
 \toprule
 \textbf{Dataset} & \textbf{Sample Snippet}  \\
 \midrule
StackOverflow & \emph{What software do you use when designing classes and their relationship, or just pen and paper?} \\
Brunet 2014 & \emph{Looks great Patrik Since this is general purpose does it belong in util Or does that introduce an unwanted dependency on dispatch} \\
SATD & \emph{// TODO: allow user to request the system or no parent}\\
Viviani 2018 & \emph{Switching the default will make all of those tutorials and chunks of code fail with routing errors, and ``the RFC says X" doesn't seem like anywhere near a good enough reason to do that.} \\
Shakiba 2016 & \emph{Move saveCallback and loadCallback to RequestProcessor class } \\
 \bottomrule
 \end{tabular}
 \end{table}

\subsection{Transfer Learning with ULMFiT}
\label{sec:ulmfit}
Early results reported by Robbes and Janes \cite{Robbes:2019} suggested recent work on ULMFiT (Universal Language Model Fine-Tuning, \cite{Howard:2018}) might work well for transfer learning in NLP for the software domain. They applied it to the task of sentiment analysis. 

 ULMFiT uses a three layer bi-LSTM (long short-term memory) architecture. It supports transfer learning for NLP tasks without having to train a new model from the beginning. ULMFiT uses novel NLP techniques like discriminative fine tuning, gradual unfreezing and slanted triangular learning rates which makes it state-of-the-art \cite{Howard:2018}. 
 ULMFiT comes pre-trained using data on a Wikipedia dataset. Wikipedia lacks software specificity.
 Our aim using ULMFiT was to observe the behavior of the pre-trained ULMFiT model (AWD-LSTM) when we train its last neural network layer with the Stack Overflow design discussion data. The StackOverflow data fine-tunes the contextual layers of the model.
 We combined the Stack Overflow data along with the four design-specific datasets for training. 

For our ULMFiT deep learning (DL) approach we used the DL practice \cite{James2013} of a 60\%-20\%-20\% train-validate-test ratio, where the test set is held back from training and validation. We ran this 3 times and noticed only trivial changes in the results for any of the runs. We report the 3rd value.

We had two reasons for this choice. First, the anticipated training costs and data size are much larger in deep network models. Second, we wanted to test the claimed capability of ULMFiT \cite{Howard:2018} to performs well on small sample sizes (60\% of our dataset). Unsurprisingly, adding more data points increases performance, at the cost of overfitting (detailed results can be found in our replication package).

Fine-tuning ULMFiT incorporates internal error assessment using a validation and train loss technique to determine the optimal trained model by constantly observing the difference between train loss and validation loss. This informs model selection, which is then tested against the held-back test set.

Training ULMFiT involves first, model pre-training using the AWD-LSTM state of the art language modelling technique; second, fine tuning the learning rate of the language model to get the optimal value of learning rate. This can be done per layer of the neural network.

\begin{figure}[hbt]
\centering
  \includegraphics[width=0.35\textwidth]{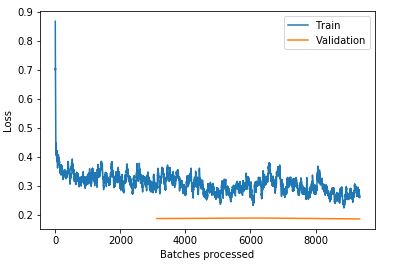}
  \caption{Loss plot after 3 epochs}
  \label{fig:Lossplot}
\end{figure}

In the third stage, train the language classification model on top of a pre-trained language learner model. The training data remains StackOverflow, but now in the supervised, design-labeled context. Finally, we again fine-tune our trained text classifier learner to find an optimal learning rate to gain a good balance between overfitting and underfitting the model. For this experiment, our optimal learning rate was 1e-2. 
After performing the above steps by combining the StackOverflow dataset---questions tagged design/non-design---and the Brunet2014 dataset, AUC was approximately 93\% during the training phase (i.e., within sample performance).
To ensure the model is not overfitting or underfitting, we plot the recorded train and validation losses. 
We make sure that the train and validation losses are close to each other. 
Fig. \ref{fig:Lossplot} is the result after 3 epochs of model training with the learning rate of 1e-2, which is the optimal learning rate for this model. 
We see that the training loss is close to the validation loss (0.18 vs 0.22). 
This suggests the trained model is performing well. 

 \begin{figure*}[hbt]
  \begin{subfigure}{.5\textwidth}
  \centering
  \includegraphics[width=.8\linewidth]{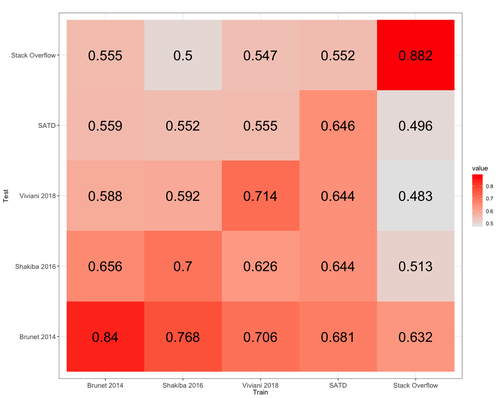}
  \caption{NewBest results.}
  \label{fig:sfig1}
\end{subfigure}
  \begin{subfigure}{.5\textwidth}
  \centering
  \includegraphics[width=.8\linewidth]{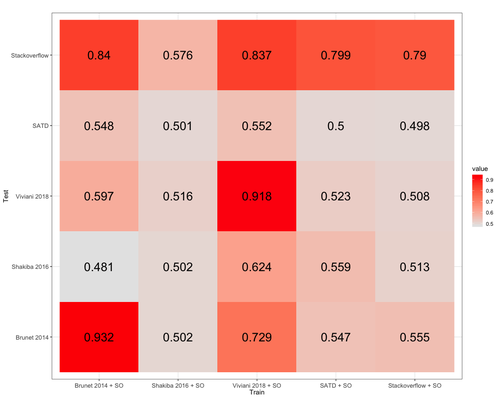}
  \caption{ULMFiT results. }
  \label{fig:sfig2}
\end{subfigure}
\caption{Cross-dataset design mining. Numbers: AUC. Read these plots as ``the model trained on the Dataset  on the X axis has AUC \emph{value} Tested On Dataset on the Y Axis". Higher intensity = better score.}
\label{fig:cross-results}
\end{figure*}

Our current approach is trained using a LSTM Neural Network. This indicates there is also scope for fine-tuning several layers of the neural network in order to gain better performance on predictions. 

\subsection{Results}
Results are summarized in the heat maps shown in Fig.  \ref{fig:cross-results}. More intense color is better. On the left, Fig. \ref{fig:sfig1} shows the results for the NewBest protocol (SVM with word embeddings). On the right, Fig \ref{fig:sfig2} shows the equivalent for the ULMFiT approach. Our replication package includes complete results including confidence intervals and tests of significance.

The main challenge for conclusion stability with design mining datasets is that it is hard to normalize natural language text. 
This means while two datasets might reasonably be said to deal with design, one might have chat-like colloquial sentences, while the other has terse, template-driven comments.  We illustrate this difference with the example discussions shown in Table \ref{tbl:examples}.
In comparing to other datasets the comparison should still be over reasonably similar `apples'. As Table \ref{tbl:newdatasets} shows, there is some variance in all five datasets, with the type of discussion artifact, source projects, and linguistic characteristics differing. However, despite these differences Table \ref{tbl:examples} suggests there should be broad similarities: e.g., concepts such as \texttt{Class} or \texttt{User}, or ideas like moving functionality to different locations. Intuitively, we suggest the notion of transfer ought to work to some extent on these datasets: they are not completely different.

For the NewBest approach, the diagonal starting bottom-left captures the \emph{within-dataset} performance, which as expected, is better than the cross-dataset AUC scores. Secondly, all models performed best on the Brunet test dataset (bottom row). This is because in building the NewBest classifier, we evaluated our protocol choices against the Brunet dataset. This shows how tightly coupled protocol choices and conclusion stability are.

It also seems to be the case that results are poorer for datasets that are more removed from each other: using pull requests (Viviani and Brunet) does little better than random for StackOverflow and code comments (SATD).

For the ULMFiT results in Fig. \ref{fig:sfig2}, we can see the benefit of training with the StackOverflow dataset plus the fine-tuning (i.e., each tic on the x-axis reflects the fine-tuning in addition to StackOverflow training). This is shown by the good results in the top row, which are essentially within-dataset results. Other results, however, are poor, and particularly when compared to the NewBest results (which is also quicker to train). A Kruskal-Wallis test of significance shows no differences between the two learners. We therefore do not see much benefit from the way we have applied ULMFiT for cross-dataset classification of design. 

We ran 10-fold cross-validation on ULMFiT as well, and report those results in the replication package. For the transfer learning research objective, ten fold validation does not make a significant difference in AUC. Transfer performance on Brunet2014 data only changes from 0.502 to 0.588, even with the additional data. Thus, our conclusion that ULMFiT is not a significant improvement on the transfer learning problem remains the same.

Although the design discussions were in natural language, there were many words that were unique to the software/open source domain. We observed that the more vocabulary we feed to our ULMFiT model, the more it knows, the more it gets tuned on the language modelling, the better it performed. Thus we see great benefit in increasing the software-specific, dataset agnostic data we train ULMFiT.

\section{Discussion}
\label{sect:discussion_and_threats_to_validity}

We discuss the implications of our results for future design mining studies, discuss the ways to improve future studies and account for researcher degrees of freedom. We begin with threats to validity for this work.

\subsection{Threats to Validity: Bad Analytics Smells}
We use the concept of bad analytics smells from Menzies and Sheppherd \cite{Menzies2019}. In that paper, the authors introduce a succinct list of twelve potential study design flaws in analytics research, and suggest some mitigations. Here, we list the smells this paper might emit, and ignore the ones we believe we have dealt with or do not apply. 

\begin{itemize}
  \item \emph{Using suspect data}: we rely extensively on \cite{Brunet2014}. However, we extend that with StackOverflow and also use datasets from other papers.
  \item \emph{Low power}: ultimately, the design mining data relies on a limited set of labeled data (or makes the possibly invalid assumption that the tagging in Stack Overflow reflects real design). Ultimately, the best solution may well be to expand the set of useful, labeled design data.  
  \item \emph{No data visualizations}: we present a limited set of visualizations, but do explore the imbalance issue, and give some examples of the data we rely on.
  \item \emph{Not tuning}: we conducted grid search on hyper-parameters \cite{xia18}. We extensively tune the ULMFiT model. We did not observe major improvements from grid search, however.
  \item \emph{Not justifying choice of algorithm}: we use the standard machine learning approaches. It is possible but unlikely that some other learner could improve results.
\end{itemize}

One other limitation is that expecting design mining classifiers trained on one dataset to transfer to totally different datasets is improbable. That is to say, machine learning is innately tightly coupled and optimized for a particular dataset. However, as Table \ref{tbl:newdatasets} shows, these design datasets are not so different. Furthermore, the literature on design patterns, and our intuition from consulting with many different projects, supports the notion that some high-order design structure crosses project boundaries.

\subsection{Improving Design Mining}
Recall that our first research objective, \textbf{RO1}, was whether it was possible to accurately label a given natural language discussion as pertaining to software design. We showed that with a judicious use of analysis choices, both of our classifiers (NewBest and ULMFiT) outperformed previous studies, and a naive baseline classifier in this task. 
However, with regard to \textbf{RO2}, the conclusion stability of our approach was low. 
Our accuracy suffers once we apply that same classifier on entirely different discussion sets. 

The problem lies in the overfitting---biasing---of the classifier to a particular dataset, given our study begins by making analysis choices that improve local (within-dataset) accuracy. This is because researchers are implicitly or explicitly conditioning on the dataset they analyze. The result is poor conclusion stability. Can we do better? 
Our results show that biased learners are a problem; applying a classifier trained on one dataset to a different dataset had very poor performance. Design discussions can change venue (issue trackers vs chat vs StackOverflow), are highly dependent on \emph{who} is communicating, and have different conceptualizations of software design. For example, if we train the document embedding on StackOverflow, and apply it to the Brunet dataset of issue discussions, our accuracy is 0.48, well below the simple ZeroR classifier which would predict the majority class, and achieve around 86\%. If we add the Brunet data to the document embedding, the accuracy doubles. However, this requires one to add and retrain the embedding each time, which is time-consuming. 

What is the generative model that leads to design discussion features? Viviani et al. \cite{Viviani2019} have begun promising work in this area by looking at higher order features such as the location of the discussion or the amount of comments on an issue. However, we likely need to look at even more robust features such as social interactions and other contextual cues. 

\subsection{The Role of Researcher Degrees of Freedom}
Researcher degrees of freedom (RDOF) \cite{Gelman:2013aa,Gelman:2012aa} refers to the multiple, equally probable analysis paths present in any research study, any of which might lead to a \emph{significant} result.  
Failure to account for researcher degrees of freedom directly impacts conclusion stability and overall practical relevance of the work, as shown in papers such as Di Nucci et al. \cite{DiNucci2018} and Hill et al. \cite{Hill2012}.
For example, for many decisions in mining studies like this one, there are competing views on when and how they should be used, multiple possible pre-processing choices, and several ways to interpret results. 
Indeed, the many combinations possible in Fig. \ref{fig:extensions} is actually over-simplified, given the actual number of choices we encountered. 
Furthermore, the existence of some choices may not be apparent to someone not deeply skilled in these types of studies. 

One possible approach is to use toolkits with intelligently tuned parameter choices. 
Hyper-parameter tuning is one such example of applying machine learning to the problem of machine learning, and research is promising \cite{xia18}. 
Clearly one particular analysis path will not apply broadly to \emph{all} software projects.
What we should aim for, however, is to outline the edges of where and more importantly, why these differences exist. 
We think there are three major steps to take to help solve the RDOF question, and improve conclusion stability.

\begin{enumerate}
  \item Use protocol maps or other graphical models to clearly outline the degrees of freedom, and chosen paths. Improve study reporting in general. There are lessons to be learned from scientific workflow software already well-developed in, for example, high-energy physics.
  \item For confirmatory studies, pre-registered hypotheses and protocols, like in medicine, make it clear what conclusions are valid, and which might be conditioned on the observed results.
  \item Improve understanding of the concept of RDOF. Develop tools that can automatically generate protocol maps based on common data science pathways. For example, we could apply our existing strengths in software slicing to analyze available parameter choices and dependencies in machine learning frameworks.
\end{enumerate}

\subsection{Future Work}

Like Shakiba et al. \cite{Shakiba2016} and Viviani \cite{Viviani2018b}, we envision a design tagging tool that can be applied broadly to all design discussions. This would be a necessary first step in automatically analyzing design decisions and recommending alternatives or improvements. To get to that point, the community needs to increase the amount of data available for these sorts of mining tasks. StackOverflow, as we demonstrate, is one potentially rich source for labeled data. More importantly, a better understanding of the nature of design discussions is needed. Expanding on qualitative studies such as Viviani et al. \cite{Viviani2018b} or those surveyed in van Vliet and Tang \cite{vanVliet2016} are the likely way forward, as opposed to blindly mining relatively small samples of software artifacts. 

\section{Conclusion}
\label{sect:conclusion}

This paper has shown new state of the art results in classifying software design discussions, with a maximum AUC score of 0.84 (using the combination of support vector machines, rebalancing, and word embedding vectorization). 
We also showed that conclusion stability for design mining remains poor. 
When we applied these best-in-class results to totally different, yet plausibly design-related datasets, our classifier achieved barely better than a naive ZeroR baseline. This was also true for state-of-the-art approaches to NLP transfer learning using the ULMFiT \cite{Howard:2018} approach.

We suggest that future studies in design mining focus more on pre-registered protocols for confirmatory approaches, and careful documentation for exploratory studies. 
Finally, conclusion stability will be improved by more labeled data, and more understanding of the differences in design discussions across projects.

\section{Acknowledgements}
We would like to thank the authors of previous work on design mining, in particular Jo\~{a}o Brunet, Giovanni Viviani, and Robert Green, for sharing their code, dataset and replication packages.

\bibliographystyle{IEEEtran}
\bibliography{abbr,saner20}

\begin{thebibliography}{10}
\providecommand{\url}[1]{#1}
\csname url@samestyle\endcsname
\providecommand{\newblock}{\relax}
\providecommand{\bibinfo}[2]{#2}
\providecommand{\BIBentrySTDinterwordspacing}{\spaceskip=0pt\relax}
\providecommand{\BIBentryALTinterwordstretchfactor}{4}
\providecommand{\BIBentryALTinterwordspacing}{\spaceskip=\fontdimen2\font plus
\BIBentryALTinterwordstretchfactor\fontdimen3\font minus
  \fontdimen4\font\relax}
\providecommand{\BIBforeignlanguage}[2]{{%
\expandafter\ifx\csname l@#1\endcsname\relax
\typeout{** WARNING: IEEEtran.bst: No hyphenation pattern has been}%
\typeout{** loaded for the language `#1'. Using the pattern for}%
\typeout{** the default language instead.}%
\else
\language=\csname l@#1\endcsname
\fi
#2}}
\providecommand{\BIBdecl}{\relax}
\BIBdecl

\bibitem{Kazman2016}
R.~Kazman and H.~Cervantes, \emph{Designing Software Architectures: A Practical
  Approach}, ser. SEI Series in Software Engineering.\hskip 1em plus 0.5em
  minus 0.4em\relax Addison-Wesley, 2016.

\bibitem{woods16}
E.~Woods, ``Software architecture in a changing world,'' \emph{IEEE Software},
  vol.~33, no.~6, pp. 94--97, Nov 2016.

\bibitem{Software-Engineering-Institute:2010aa}
{Software Engineering Institute}, ``What is your definition of software
  architecture?'' Software Engineering Institute, Fact Sheet, 2010.

\bibitem{Shahbazian2018}
\BIBentryALTinterwordspacing
A.~Shahbazian, Y.~K. Lee, D.~Le, Y.~Brun, and N.~Medvidovic, ``Recovering
  architectural design decisions,'' in \emph{Proceedings of the IEEE
  International Conference on Software Architecture}.\hskip 1em plus 0.5em
  minus 0.4em\relax {IEEE}, apr 2018. [Online]. Available:
  \url{https://doi.org/10.1109/icsa.2018.00019}
\BIBentrySTDinterwordspacing

\bibitem{Brunet2014}
J.~Brunet, G.~C. Murphy, R.~Terra, J.~Figueiredo, and D.~Serey, ``Do developers
  discuss design?'' in \emph{Working Conference on Mining Software
  Repositories}, Hyderabad, India, September 2014.

\bibitem{Shakiba2016}
\BIBentryALTinterwordspacing
A.~Shakiba, R.~Green, and R.~Dyer, ``{FourD}: do developers discuss design?
  revisited,'' in \emph{Proceedings of the 2nd International Workshop on
  Software Analytics - {SWAN} 2016}.\hskip 1em plus 0.5em minus 0.4em\relax
  {ACM} Press, 2016. [Online]. Available:
  \url{https://doi.org/10.1145/2989238.2989244}
\BIBentrySTDinterwordspacing

\bibitem{Alkadhi2017}
R.~Alkadhi, T.~Lata, E.~Guzmany, and B.~Bruegge, ``Rationale in development
  chat messages: An exploratory study,'' in \emph{Proceedings of the
  International Working Conference on Mining Software Repositories}, may 2017.

\bibitem{Motta2018}
\BIBentryALTinterwordspacing
T.~O. Motta, R.~R. {Gomes e Souza}, and C.~Sant'Anna, ``Characterizing
  architectural information in commit messages,'' in \emph{Proceedings of the
  Brazilian Symposium on Software Engineering}.\hskip 1em plus 0.5em minus
  0.4em\relax {ACM} Press, 2018. [Online]. Available:
  \url{https://doi.org/10.1145/3266237.3266260}
\BIBentrySTDinterwordspacing

\bibitem{Zanaty2018}
\BIBentryALTinterwordspacing
F.~E. Zanaty, T.~Hirao, S.~McIntosh, A.~Ihara, and K.~Matsumoto, ``An empirical
  study of design discussions in code review,'' in \emph{Proceedings of the
  International Symposium on Empirical Software Engineering and
  Measurement}.\hskip 1em plus 0.5em minus 0.4em\relax {ACM} Press, 2018.
  [Online]. Available: \url{https://doi.org/10.1145/3239235.3239525}
\BIBentrySTDinterwordspacing

\bibitem{Viviani2018b}
G.~Viviani, C.~Janik-Jones, M.~Famelis, and G.~C. Murphy, ``The structure of
  software design discussions,'' in \emph{Proceedings of the International
  Workshop on Cooperative and Human Aspects of Software Engineering}.\hskip 1em
  plus 0.5em minus 0.4em\relax {ACM} Press, 2018.

\bibitem{Viviani2019}
G.~{Viviani}, M.~{Famelis}, X.~{Xia}, C.~{Janik-Jones}, and G.~C. {Murphy},
  ``Locating latent design information in developer discussions: A study on
  pull requests,'' \emph{IEEE Transactions on Software Engineering}, pp. 1--1,
  2019.

\bibitem{Oliveira-Neto:2019aa}
\BIBentryALTinterwordspacing
F.~G. de~Oliveira~Neto, R.~Torkar, R.~Feldt, L.~Gren, and C.~A. Furia,
  ``Evolution of statistical analysis in empirical software engineering
  research: Current state and steps forward,'' \emph{Journal of Systems and
  Software}, 2019. [Online]. Available: \url{https://arxiv.org/abs/1706.00933}
\BIBentrySTDinterwordspacing

\bibitem{Menzies2012}
\BIBentryALTinterwordspacing
T.~Menzies and M.~Shepperd, ``Special issue on repeatable results in software
  engineering prediction,'' \emph{Empirical Software Engineering}, vol.~17,
  no.~1, pp. 1--17, Feb 2012. [Online]. Available:
  \url{https://doi.org/10.1007/s10664-011-9193-5}
\BIBentrySTDinterwordspacing

\bibitem{Gmez2014}
\BIBentryALTinterwordspacing
O.~S. G{\'{o}}mez, N.~Juristo, and S.~Vegas, ``Understanding replication of
  experiments in software engineering: A classification,'' \emph{Information
  and Software Technology}, vol.~56, no.~8, pp. 1033--1048, aug 2014. [Online].
  Available: \url{https://doi.org/10.1016/j.infsof.2014.04.004}
\BIBentrySTDinterwordspacing

\bibitem{Siegmund2015}
\BIBentryALTinterwordspacing
J.~Siegmund, N.~Siegmund, and S.~Apel, ``Views on internal and external
  validity in empirical software engineering,'' in \emph{2015 {IEEE}/{ACM} 37th
  {IEEE} International Conference on Software Engineering}.\hskip 1em plus
  0.5em minus 0.4em\relax {IEEE}, May 2015. [Online]. Available:
  \url{https://doi.org/10.1109/icse.2015.24}
\BIBentrySTDinterwordspacing

\bibitem{Howard:2018}
J.~Howard and S.~Ruder, ``Universal language model fine-tuning for text
  classification,'' in \emph{Annual Meeting of the Association for
  Computational Linguistics}, 2018.

\bibitem{Zimmermann:2009}
T.~Zimmermann, N.~Nagappan, H.~Gall, E.~Giger, and B.~Murphy, ``Cross-project
  defect prediction: A large scale experiment on data vs. domain vs. process,''
  in \emph{Proceedings of the European Software Engineering Conference/ACM
  SIGSOFT International Symposium on Foundations of Software Engineering},
  2009, pp. 91--100.

\bibitem{Pan:2010}
\BIBentryALTinterwordspacing
S.~J. Pan and Q.~Yang, ``A survey on transfer learning,'' \emph{IEEE
  Transactions on Knowledge and Data Engineering}, vol.~22, no.~10, pp.
  1345--1359, Oct. 2010. [Online]. Available:
  \url{https://doi.org/10.1109/TKDE.2009.191}
\BIBentrySTDinterwordspacing

\bibitem{Sharma:2019aa}
T.~Sharma, V.~Efstathiou, P.~Louridas, and D.~Spinellis, ``On the feasibility
  of transfer-learning code smells using deep learning,'' arXiv, Tech. Rep.
  1904.03031v2, 2019.

\bibitem{Krishna:2016:TMA:2970276.2970339}
\BIBentryALTinterwordspacing
R.~Krishna, T.~Menzies, and W.~Fu, ``Too much automation? the bellwether effect
  and its implications for transfer learning,'' in \emph{International
  Conference on Automated Software Engineering}, 2016, pp. 122--131. [Online].
  Available: \url{http://doi.acm.org/10.1145/2970276.2970339}
\BIBentrySTDinterwordspacing

\bibitem{Robbes:2019}
\BIBentryALTinterwordspacing
R.~Robbes and A.~Janes, ``Leveraging small software engineering data sets with
  pre-trained neural networks,'' in \emph{International Conference on Software
  Engineering: New Ideas and Emerging Results}, ser. ICSE-NIER '19, 2019, pp.
  29--32. [Online]. Available:
  \url{https://doi.org/10.1109/ICSE-NIER.2019.00016}
\BIBentrySTDinterwordspacing

\bibitem{hindle11msr}
\BIBentryALTinterwordspacing
A.~Hindle, N.~Ernst, M.~W. Godfrey, and J.~Mylopoulos, ``{Automated topic
  naming to support cross-project analysis of software maintenance
  activities},'' in \emph{MSR}, Honolulu, 2011, pp. 1--10. [Online]. Available:
  \url{http://portal.acm.org/citation.cfm?id=1985441.1985466&coll=DL&dl=GUIDE&CFID=174423731&CFTOKEN=59920133}
\BIBentrySTDinterwordspacing

\bibitem{HindleBZN15}
\BIBentryALTinterwordspacing
A.~Hindle, C.~Bird, T.~Zimmermann, and N.~Nagappan, ``Do topics make sense to
  managers and developers?'' \emph{Empirical Software Engineering}, vol.~20,
  no.~2, pp. 479--515, 2015. [Online]. Available:
  \url{https://doi.org/10.1007/s10664-014-9312-1}
\BIBentrySTDinterwordspacing

\bibitem{Pickard1998}
\BIBentryALTinterwordspacing
L.~M. Pickard, B.~A. Kitchenham, and P.~W. Jones, ``Combining empirical results
  in software engineering,'' \emph{Information and Software Technology},
  vol.~40, no.~14, pp. 811 -- 821, 1998. [Online]. Available:
  \url{http://www.sciencedirect.com/science/article/pii/S0950584998001013}
\BIBentrySTDinterwordspacing

\bibitem{Alkadhi2018}
\BIBentryALTinterwordspacing
R.~Alkadhi, M.~Nonnenmacher, E.~Guzman, and B.~Bruegge, ``How do developers
  discuss rationale?'' in \emph{Proceedings of the IEEE International
  Conference on Software Analysis, Evolution, and Reengineering}.\hskip 1em
  plus 0.5em minus 0.4em\relax {IEEE}, mar 2018. [Online]. Available:
  \url{https://doi.org/10.1109/saner.2018.8330223}
\BIBentrySTDinterwordspacing

\bibitem{Maldonado2017}
\BIBentryALTinterwordspacing
E.~da~Silva~Maldonado, E.~Shihab, and N.~Tsantalis, ``Using natural language
  processing to automatically detect self-admitted technical debt,'' \emph{IEEE
  Transactions on Software Engineering}, vol.~43, no.~11, pp. 1044--1062, nov
  2017. [Online]. Available: \url{https://doi.org/10.1109/tse.2017.2654244}
\BIBentrySTDinterwordspacing

\bibitem{Viviani2018}
G.~Viviani, C.~Janik-Jones, M.~Famelis, X.~Xia, and G.~C. Murphy, ``What design
  topics do developers discuss?'' in \emph{Proceedings of the IEEE
  International Conference on Program Comprehension}, 2018.

\bibitem{Aranda_2009}
\BIBentryALTinterwordspacing
J.~Aranda and G.~Venolia, ``The secret life of bugs: Going past the errors and
  omissions in software repositories,'' \emph{Proceedings of the ACM/IEEE
  International Conference on Software Engineering}, 2009. [Online]. Available:
  \url{http://dx.doi.org/10.1109/ICSE.2009.5070530}
\BIBentrySTDinterwordspacing

\bibitem{Ernst:2012wf}
N.~Ernst and G.~C. Murphy, ``{Case Studies in Just-In-Time Requirements
  Analysis},'' in \emph{Empirical Requirements Engineering Workshop at RE},
  Chicago, Sep. 2012, pp. 1--8.

\bibitem{Kitchenham2019}
\BIBentryALTinterwordspacing
B.~Kitchenham, L.~Madeyski, and P.~Brereton, ``Meta-analysis for families of
  experiments in software engineering: a systematic review and reproducibility
  and validity assessment,'' \emph{Empirical Software Engineering}, Jul 2019.
  [Online]. Available: \url{https://doi.org/10.1007/s10664-019-09747-0}
\BIBentrySTDinterwordspacing

\bibitem{Storey:2019aa}
M.-A. Storey, C.~Williams, N.~A. Ernst, A.~Zagalsky, and E.~Kalliamvakou,
  ``Methodology matters: How we study socio-technical aspects in software
  engineering,'' arXiv, Tech. Rep. arXiv:1905.12841, 2019.

\bibitem{Shepperd2018}
M.~Shepperd, ``Replication studies considered harmful,'' in \emph{Companion of
  the International Conference on Software Engineering}, 2018.

\bibitem{mikolov2018advances}
T.~Mikolov, E.~Grave, P.~Bojanowski, C.~Puhrsch, and A.~Joulin, ``Advances in
  pre-training distributed word representations,'' in \emph{Proceedings of the
  International Conference on Language Resources and Evaluation (LREC 2018)},
  2018.

\bibitem{pennington2014glove}
\BIBentryALTinterwordspacing
J.~Pennington, R.~Socher, and C.~D. Manning, ``Glove: Global vectors for word
  representation,'' in \emph{Empirical Methods in Natural Language Processing
  (EMNLP)}, 2014, pp. 1532--1543. [Online]. Available:
  \url{http://www.aclweb.org/anthology/D14-1162}
\BIBentrySTDinterwordspacing

\bibitem{Efstathiou:2018aa}
\BIBentryALTinterwordspacing
V.~Efstathiou, C.~Chatzilenas, and D.~Spinellis, ``Word embeddings for the
  software engineering domain,'' in \emph{Proceedings of the International
  Working Conference on Mining Software Repositories}, 2018. [Online].
  Available: \url{https://github.com/vefstathiou/SO_word2vec}
\BIBentrySTDinterwordspacing

\bibitem{doc2vec14}
\BIBentryALTinterwordspacing
Q.~Le and T.~Mikolov, ``Distributed representations of sentences and
  documents,'' in \emph{Proceedings of the International Conference on Machine
  Learning}, 2014. [Online]. Available:
  \url{https://cs.stanford.edu/~quocle/paragraph_vector.pdf}
\BIBentrySTDinterwordspacing

\bibitem{BaltesDT008}
S.~Baltes, L.~Dumani, C.~Treude, and S.~Diehl, ``Sotorrent: reconstructing and
  analyzing the evolution of stack overflow posts,'' in \emph{Proceedings of
  the International Working Conference on Mining Software Repositories}, 2018,
  pp. 319--330.

\bibitem{Chawla:2002aa}
N.~Chawla, K.~Bowyer, L.~Hall, and W.~P. Kegelmeyer, ``Smote: Synthetic
  minority over-sampling technique,'' \emph{Journal of Artificial Intelligence
  Research}, vol.~16, pp. 321--357, 2002.

\bibitem{James2013}
\BIBentryALTinterwordspacing
G.~James, D.~Witten, T.~Hastie, and R.~Tibshirani, \emph{An Introduction to
  Statistical Learning: with Applications in R}.\hskip 1em plus 0.5em minus
  0.4em\relax Springer, 2013. [Online]. Available:
  \url{http://www-bcf.usc.edu/~gareth/ISL/getbook.html}
\BIBentrySTDinterwordspacing

\bibitem{Menzies2019}
\BIBentryALTinterwordspacing
T.~Menzies and M.~Shepperd, ````{B}ad smells'' in software analytics papers,''
  \emph{Information and Software Technology}, vol. 112, pp. 35--47, Aug. 2019.
  [Online]. Available: \url{https://doi.org/10.1016/j.infsof.2019.04.005}
\BIBentrySTDinterwordspacing

\bibitem{xia18}
\BIBentryALTinterwordspacing
T.~Xia, R.~Krishna, J.~Chen, G.~Mathew, X.~Shen, and T.~Menzies,
  ``Hyperparameter optimization for effort estimation,'' ArXiv, Tech. Rep.,
  2018. [Online]. Available: \url{http://arxiv.org/abs/1805.00336}
\BIBentrySTDinterwordspacing

\bibitem{Gelman:2013aa}
A.~Gelman and E.~Loken, ``The garden of forking paths: Why multiple comparisons
  can be a problem, even when there is no ``fishing expedition'' or
  ``p-hacking'' and the research hypothesis was posited ahead of time,''
  Colombia University, Tech. Rep., 2013.

\bibitem{Gelman:2012aa}
A.~Gelman, J.~Hill, and M.~Yajima, ``Why we (usually) don't have to worry about
  multiple comparisons,'' \emph{Journal of Research on Educational
  Effectiveness}, vol.~5, pp. 189--211, 2012.

\bibitem{DiNucci2018}
\BIBentryALTinterwordspacing
D.~D. Nucci, F.~Palomba, D.~A. Tamburri, A.~Serebrenik, and A.~D. Lucia,
  ``Detecting code smells using machine learning techniques: Are we there
  yet?'' in \emph{International Conference on Software Analysis, Evolution and
  Reengineering ({SANER})}, Mar. 2018. [Online]. Available:
  \url{https://doi.org/10.1109/saner.2018.8330266}
\BIBentrySTDinterwordspacing

\bibitem{Hill2012}
\BIBentryALTinterwordspacing
E.~Hill, S.~Rao, and A.~Kak, ``On the use of stemming for concern location and
  bug localization in java,'' in \emph{International Working Conference on
  Source Code Analysis and Manipulation}.\hskip 1em plus 0.5em minus
  0.4em\relax {IEEE}, Sep. 2012. [Online]. Available:
  \url{https://doi.org/10.1109/scam.2012.29}
\BIBentrySTDinterwordspacing

\bibitem{vanVliet2016}
H.~van Vliet and A.~Tang, ``Decision making in software architecture,''
  \emph{Journal of Systems and Software}, vol. 117, pp. 638--644, jul 2016.

\end{thebibliography}

\end{document}